\documentclass[final, numberedheadings]{aipproc}

\usepackage{amsmath, amssymb, latexsym}

\layoutstyle{6x9}

\newcommand{\fstat}{$\mathcal{F}$-statistic}

\newcommand{\snd}{\textit{Slice \& Dice}}

\begin{document}

\title{Slice \& Dice: Identifying and Removing Bright Galactic
Binaries from LISA Data}



\author{Louis J. Rubbo}{
  address={Center for Gravitational Wave Physics, Pennsylvania State
  University, University Park, PA 16802}
}

\author{Neil J. Cornish}{
  address={Department of Physics, Montana State University - Bozeman,
  Bozeman, MT 59717}
}

\author{Ronald W. Hellings}{
  address={Department of Physics, Montana State University - Bozeman,
  Bozeman, MT 59717}
}



\begin{abstract}
Here we describe a hierarchal and iterative data analysis algorithm
used for searching, characterizing, and removing bright, monochromatic
binaries from the Laser Interferometer Space Antenna (LISA) data
streams.  The algorithm uses the \fstat\ to provide an initial
solution for individual bright sources, followed by an iterative least
squares fitting for all the bright sources.  Using the above
algorithm, referred to as \snd, we demonstrate the removal of
multiple, correlated galactic binaries from simulated LISA data.
Initial results indicate that \snd\ may be a useful tool for analyzing
the forthcoming LISA data.
\end{abstract}

\maketitle


\section{INTRODUCTION} \label{sec:intro}

Inside the Milky Way galaxy there is a plethora of binaries whose
emitted gravitational waves have a frequency inside the Laser
Interferometer Space Antenna (LISA) band~\cite{Evans:1987, Hils:1990,
Nelemans:2001, Benacquista:2004}.  For the vast majority of these
sources, the measured signals will be buried in the instrument noise.
However, thousands of sources will have signals strong enough to be
detected above the instrument noise at low frequencies ($0.1 \lesssim
f \lesssim$~3~mHz).  The superposition of these sources will form a
confusion limited background below which individual systems cannot be
distinguished from the collective population.  This confusion
background will act as an additional noise component for the detector.

Lying above the confusion background will be a number of individual
sources whose signals are stronger than the local rms value of the
background~\cite{Timpano:2006}.  Due to their relative brightness,
they will be resolvable within the LISA data streams.  Additionally,
at higher frequencies, where the confusion background drops below the
detector noise, individual sources will also be resolvable.  Due to
the large orbital periods and low chirp masses associated with
galactic binaries, radiation reaction effects will not drive the
binaries to coalescence during the mission lifetime, so their signals
will be ever present in the detector output.  It is these galactic
binaries that are prime targets for data analysis techniques used for
identification, characterization, and subtraction.

Even though these target signals are relatively bright, significant
correlations may still exist between them.  Consequently,
sophisticated data analysis techniques capable of resolving the
individual systems will be required.  Here we present the \snd\
algorithm which was developed to resolve the bright galactic binary
signals that lie above the confusion limited background.  \snd\ uses
the \fstat\ (a template based routine~\cite{Jaranowski:1998}) to
identify and initially characterize the bright signals.  The \fstat\
results are then used to initiate a least squares fitting procedure
which refines the parameter estimates.

The rest of this paper proceeds as follows.  In
section~\ref{sec:order} we discuss the motivation for \snd\ by
illustrating the importance of when signals are identified and
subtracted from the data.  In section~\ref{sec:snd} we describe the
\snd\ algorithm in detail and give a simple example of its use.
Finally in section~\ref{sec:conclusion} we discuss future
developments.

\section{IDENTIFICATION AND SUBTRACTION ORDER} \label{sec:order}

Template matching and least squares fitting provide a robust approach
to identifying and subtracting individual signals from LISA's
data~\cite{Hellings:2003}.  However, when dealing with multiple
signals simultaneously the order of identification and removal is
crucial.

\begin{figure}
  \includegraphics[height=0.29\textheight]{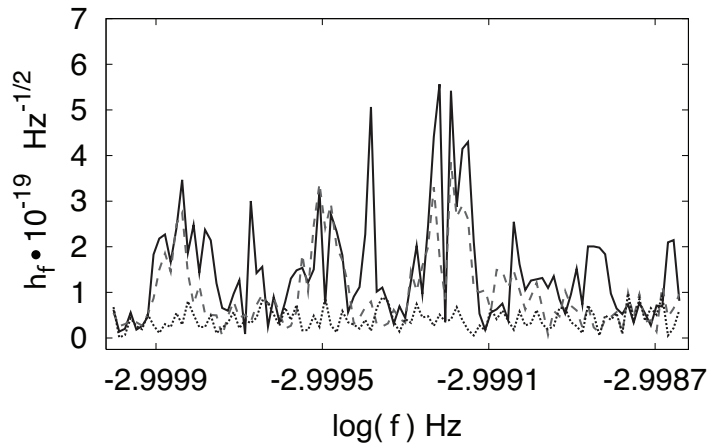}
    \caption{Results of a sequential identification, ordered by
    frequency, followed by a global removal scheme.  The solid line is
    the original data, the dashed is the residue, and the dotted is
    the instrument noise.}
  \label{fig:seqidenrem}
\end{figure}

Figure~\ref{fig:seqidenrem} shows the spectral amplitudes for a
sequential identification, ordered by frequency, followed by a global
removal using a template matching scheme on twenty signals within one
hundred frequency bins\footnote{A frequency bin $\Delta f$ is equal to
one on the observation time, $\Delta f = T^{-1}$.  For a one year
observation, which is used here, $\Delta f = 3.2 \times 10^{-8}$~Hz.}
added to Gaussian noise.  The solid line is the original data, the
dotted is the instrumental noise, and the dashed is the residue after
subtraction.  The large residue indicates that this approached failed
to accurately identify the injected signals.  The failure arose
because the algorithm treats the other comparable, bright signals as
noise while attempting to fit for one signal.

\begin{figure}
  \includegraphics[height=0.29\textheight]{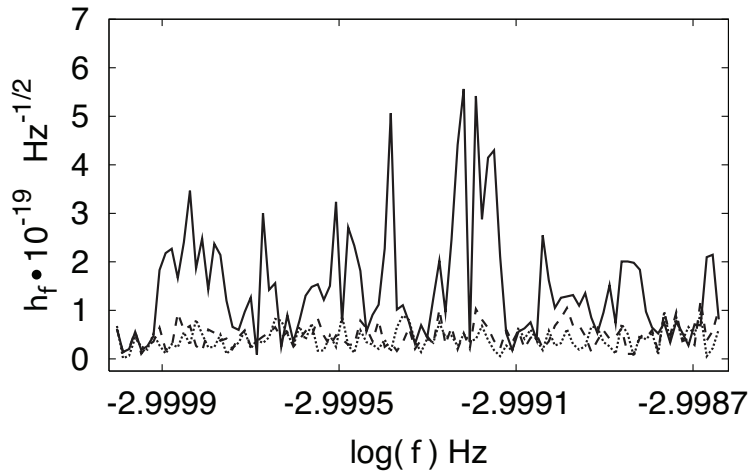}
    \caption{Results of a sequential identification and removal
    algorithm with the order is from highest SNR to the lowest.  The
    line identifications are the same as Figure~1.}
  \label{fig:seqiden}
\end{figure}

Figure~\ref{fig:seqiden} demonstrates the result of a sequential fit
and subtraction procedure where the order of fitting and removal is
based on the signal-to-noise ratios (SNRs).  Although the residue
looks reasonable, a correlation comparison of the fitted signals to
the original injections indicates that the weaker signals were not
properly identified.  As before, the interference from the other
bright signals as the algorithm attempts to fit for one signal causes
errors, which ultimately propagate through to the fits for the weaker
sources due to the intermediate subtractions.

\begin{figure}
  \includegraphics[height=0.29\textheight]{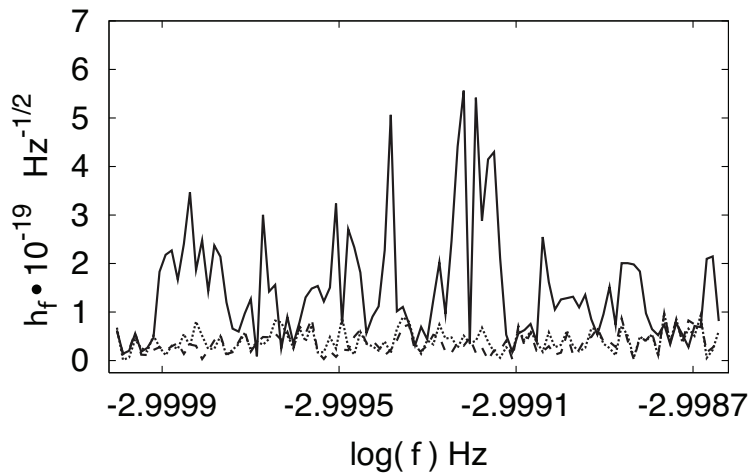}
    \caption{Results of a simultaneous identification procedure.  The
    line identifications are the same as Figure~1.}
  \label{fig:simul}
\end{figure}

Figure~\ref{fig:simul} demonstrates the result from performing a
simultaneous fitting and removal scheme using a least squares fitting
routine initiated at the true parameter values.  In this case there
are acceptable fits for each signal.  However, for this approach to
work we had to assume the initial values for the least squares
procedure.  Normally these values would be obtained from a search
algorithm, such as a template matching routine.  Unfortunately, when
the number of bright systems increases, the resulting number of
templates required to sufficiently cover the parameter space make a
template approach impractical.

\snd\ was developed to incorporate the robust attributes of template
matching and least squares fitting while circumventing the problems of
a large template bank and providing initial guesses for the least
squares routine.

\section{SLICE \& DICE ALGORITHM} \label{sec:snd}

The \snd\ algorithm is an iterative routine that incorporates multiple
LISA data streams.  Each iteration involves the following steps,
\begin{enumerate}
\item An \fstat\ (i.e. a template based) search is used to find the
brightest signal and return an initial estimate for its parameter
values.
\item The \fstat\ parameter estimates, along with previous iteration
estimates, are used to initiate a least squares fitting routine on the
original data.  The least squares routine simultaneously solves for
$i$ signals, where $i$ is the number of sources found so far.
\item The least squares fitting routine is repeated until the change
in the $\chi^{2}$ is insignificant ($<0.01$).  At each least squares
cycle the initial parameter value guesses are supplied from the
previous least squares iteration.
\item The estimated signals are subtracted from the original data
streams and this partially cleaned data is used to initiate the next
\snd\ iteration.
\end{enumerate}
These steps are repeated until each bright source (SNR $>$ 5) are
identified and removed.

As an example of \snd\ results, consider two systems with SNRs of 17
and 12, separated in frequency by $0.07 \Delta f$, in sky location by
$90^{\circ}$, and with random orientations.  The initial correlation
matrix between the injected signals is
\begin{displaymath}
  r_{1 yr} = \left( \begin{array}{cc}
  1 & 0.090 \\
  0.090 & 1
  \end{array} \right) \;.
\end{displaymath}
While their frequencies are nearly identical, these sources are
largely separated on the sky and therefore have small
cross-correlations.

\begin{figure}
  \includegraphics[height=0.29\textheight]{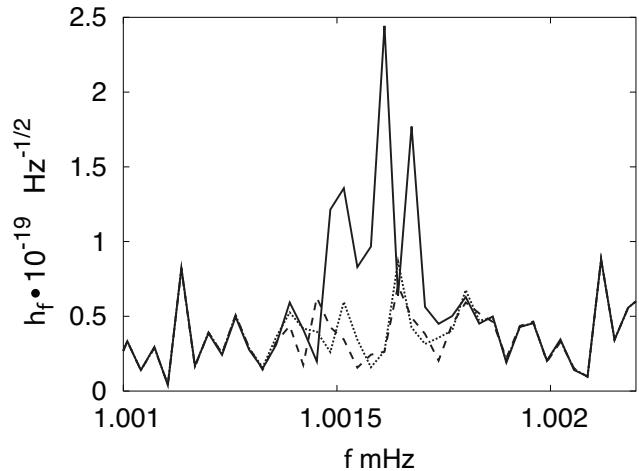}
  \caption{Results of \snd\ on two signals with nearly identical
    frequencies but largely separated on the sky.  The line
    identifications are the same as Figure~1.}
  \label{fig:snd2}
\end{figure}

Figure~\ref{fig:snd2} shows the spectral amplitudes before and after
\snd\ has been applied to the data, along with the instrument noise.
The final comparison correlation matrix between the original and the
estimated signals is
\begin{displaymath}
  r_{1 yr} = \left( \begin{array}{cc}
  0.94 & 0.11 \\
  0.06 & 0.99
  \end{array} \right) \;.
\end{displaymath}
\snd\ was able to accurately identify the two signals.

\section{FUTURE DEVELOPMENTS} \label{sec:conclusion}

At this time we are testing \snd\ in order to understand its strengths
and limitations.  We have successfully applied the algorithm on bright
source densities comparable to those shown in
Figures~\ref{fig:seqidenrem} through \ref{fig:simul} (i.e. one bright
source every five frequency bins).  However, possible parameter
degeneracies and how their effects propagate through the iterative
\snd\ routine are not fully understood.  We are currently testing how
the algorithm responds to such degeneracies, and how they may limit
its capabilities.

In addition to the above mentioned tests, we are also expanding the
scope of applicability.  \snd\ is currently only applicable on a small
segment of the LISA band.  In the future we will expand and automate
the algorithm so that it can take the full bandwidth of data.
Additionally, we would like to compare and contrast \snd's performance
to other galactic binary search procedures such as
gCLEAN~\cite{Cornish:2003}, Markov Chain Monte Carlo
methods~\cite{Cornish:2005, Umstatter:2005}, genetic
algorithms~\cite{Crowder:2006}, and maximum entropy.


\begin{theacknowledgments}
The work by LJR was supported by the Center for Gravitational Wave
Physics.  The Center for Gravitational Wave Physics is supported by
the NSF under Cooperative Agreement No. PHY 01-14375.  NJC was
supported by NASA grant NNG05GI69G.
\end{theacknowledgments}


\bibliographystyle{aipproc}
\bibliography{References}

\end{document}